\documentclass[preprint,number,12pt]{elsarticle}
\usepackage{amsmath}
\usepackage{graphicx}
\usepackage{braket}
\usepackage{dcolumn}
\usepackage{bm}
\usepackage{hyperref}
\usepackage{float}
\usepackage{xcolor}
\usepackage{subfigure}
\usepackage{ulem}
\usepackage[T1]{fontenc}
\usepackage[english]{babel}
\begin{document}
\title{Persistent revivals in a system of trapped bosonic atoms}

\author[1]{Carlos Diaz Mejia}
\ead{carlosdime@estudiantes.fisica.unam.mx}

\author[1]{Javier de la Cruz}
\ead{javierdelacruz@ciencias.unam.mx}

\author[2]{Sergio Lerma-Hern\'andez} \ead{slerma@uv.mx} 

\author[1]{Jorge G. Hirsch} 
 \ead{hirsch@nucleares.unam.mx} 

\affiliation[1]{organization={Instituto de Ciencias Nucleares, Universidad Nacional Aut\'onoma de M\'exico},addressline={ Apdo. Postal 70-543}, postcode={04510}, city={  CDMX},country={Mexico}}

\affiliation[2]{organization={Facultad de F\'isica, Universidad Veracruzana}, addressline={Campus Arco Sur, Paseo~112}, postcode={91097}, city={Xalapa},  country={Mexico}}
\date{\today}

\begin{abstract}

Dynamical signatures of quantum chaos are observed in the survival probability of different initial states, in a system of cold atoms trapped in a linear chain with site noise and open boundary conditions.  
It is shown that chaos is present in the region of small disorder, at intermediate energies.
The study is performed with  different number of sites and atoms: 7,8 and 9, but focusing on the case where the particle density is one. 
States of the occupation basis with energies in the chaotic region are evolved at long times. 

Remarkable differences in the behaviour of the survival probability are found for states with different energy-eigenbasis  participation ratio (PR). Whereas those with large PR clearly  exhibit the characteristic random-matrix correlation hole before equilibration, those with small PR present 
a marginal or even no  correlation hole which is replaced  by  revivals  lasting up to the  stage of equilibration,  suggesting  a connection with the quantum scarring phenomenon.   

\end{abstract}

\maketitle

\large

\section{\label{sec:level1}Introduction}

Quantum  chaos is a concept closely related to and deeply rooted in Random Matrix Theory (RMT), a formalism useful to describe
the correlations in  the spectrum of a quantum system with a classical correspondent that is chaotic \cite{BGS_84, Berry(1977)}. 
RMT also provides the mathematical framework to support the 
Eigenstate Thermalization Hypothesis (ETH) \cite{Murthy2019, Srednicki_94}, which  explains the thermalization of quantum many-body system by using a small set of chaotic eigenstates around a given energy, expected to have similar properties. 
In the description of chaos and thermalization the focus is on the properties of the eigenvalues and eigenvectors of the Hamiltonian. The dynamics of a given initial state  
centered in the same energy region is strongly influenced by these spectral properties, but its particular structure plays also a relevant role.
To what extent the properties of the  initial state are important for the different temporal scales in their dynamics? In which cases the dynamics is determined solely by the universal properties coming from RMT?  In what follows we address these questions for a system of trapped atoms and a set of experimentally relevant initial states.

Experiments and numerical simulations in the Interacting Aubry-Andre model (IAA), consisting of eight bosons and eight sites \cite{Lukin_2019}, have explored the presence of thermalization. The central observable was the entanglement entropy of one site with the rest at long times, for an initial Mott state with one boson per site. The study showed the presence of thermalization for weak disorder, and that, for a large disorder parameter,
the relaxation value of the entanglement entropy did not match the expected thermal average provided by the ETH. This phenomenology is attributed to many-body localization (MBL)\cite{Serbyn_2021}, where non-local correlations would not persist during the evolution of the observable.

A different manifestation of weak ergodicity breaking is the existence of persistent revivals of observables\cite{chandran2023quantum}. Typically, they occur when ETH and quantum chaos are valid. Since the discovery of this oscillatory phenomena in Rydberg chains\cite{bernien2017probing,turner2018quantum}, many experiments and theoretical\cite{choi2019emergent} studies have arisen to extend the research on quantum scars. Some examples are: Hubbard models\cite{Hudomal2020,hummel2023genuine}, spin chains \cite{dooley2020enhancing,labuhn2016tunable} and quantum simulators\cite{hummel2023genuine,su2023observation}, etc. There are still some open questions regarding why there are some states that only explore a small set of a very large Hilbert space.

In this work, our central results focus on the chaotic region, where ETH is valid, and even so, some initial states do not evolve as expected according to RMT. 
We  present a detailed study of the quantum dynamical properties of the Interacting Aubry-Andr\'e model for a set of occupation (Fock) initial states. We performed a scaling analysis from 7 bosons and 7 sites up  to 9 bosons and 9 sites, which have   Hilbert-space dimensions, $\textrm{dim}=\binom{2N-1}{N-1}$, from $1716$ up to  $24310$, and confirmed that  the  chaotic (RMT) correlations in the spectrum  remain at the same interval of energy per particle.

A classification of all the possible initial states on the site basis is developed
introducing a crowding parameter, which is correlated with the expectation value of the energy of the occupation states, averaged over 
different realizations of the Hamiltonian. It is shown that the states whose energies lie in the chaotic region have larger participation ratios (PR) in the eigenbasis than  those in the regular region, which  have PR noticeably smaller. The states with the largest PR in the chaotic region exhibit the so-called  correlation hole in their survival probability, while the states with  the smallest 
PR in the same chaotic region, have a survival probability  not showing such  correlation hole  and have lasting revivals persisting up to the equilibration  stage of their evolution.

The structure of the paper is as follows: 
in section \ref{sec:model} we introduce the Hamiltonian of the system with emphasis on 
the energy level spacing ratio to  determine the values of the disorder parameter and the energies where quantum chaos is present, 
in section \ref{sec:initial_states} the states in the occupation basis are classified according to the crowding parameter and Participation Ratio in the Hamiltonian eigenbasis.
  In section  \ref{sec:survival_probability} we analyze 
the survival probability of the occupation states, and the presence or absence of the correlation hole, which is the hallmark of chaotic dynamics. 
Conclusions are presented in section \ref{sec:conclusion}. In the Appendix some technical aspects of the evaluation of the Survival Probability are presented.   

\section{\label{sec:model}The Interacting Aubry-Andr\'e model (IAA) and spectral signatures of quantum chaos}

The  Interacting Aubry-Andr\'e model (IAA)  describes the dynamics of $N$ spin-less bosons on a lattice with $L$ sites in a one dimensional chain\cite{Paredes}. The corresponding Hamiltonian($\hbar= 1$) is
\begin{equation}
H = -J \sum_{\braket{i,j}}\hat{b}_{i}^{\dagger}\hat{b}_{j}  + \frac{U}{2} \sum_{i}\hat{n}_{i}(\hat{n}_{i}-1)
+ W \sum_i \cos(2\pi \beta i + \phi) \hat{n}_i,\label{eq:Hamiltonian}
\end{equation}
where $\hat{b}^{\dagger}_i$ and $\hat{b}_i$ are the bosonic creation and annihilation operators on the site $i$.
The first term describes the tunneling between neighboring lattice sites with  rate $J$. The second term represents the energy shift $U$ when multiple particles occupy the same site, with $\hat{n}_i=\hat{b}_i^{\dagger}\hat{b}_i$ the number of particles on site $i$. The last term introduces a site-resolved potential offset, which is created with an incommensurate lattice $\cos (2 \pi \beta i + \phi)$ of period $\beta$ lattice sites, phase $\phi$, and amplitude $W$,  known as disorder parameter.
The Hamiltonian parameters $J$, $U$,  $W$ and $\phi$ can be controlled in the experiment \cite{Lukin_2019}. In our study  the phase $\phi$ varies randomly from one realization to another in order to make clearer the spectrum correlations in the observable we considered.

For large values of the disorder parameter $W$ in  Hamiltonian (\ref{eq:Hamiltonian}) many-body localization (MBL) breaks thermalization; particles are localized and transport ceases. This localization transition is very sensitive to the value of the parameter $\beta$, which is selected as an incommensurate number, the ratio of two prime numbers $\beta=P/Q$. In the experiment it is related  with the two laser wavelengths. In this work the values $\beta=1.618$  \cite{Lukin_2019},   $U=4/(N-1)$ and $J=1/2$ are employed in the calculations. The parameter $U$ is scaled by factor $1/(N-1)$ in order to obtain the same scaling properties for all the terms in the Hamiltonian. The latter values are associated with the   presence of chaos when $W=0$ \cite{Kolovsky_2004,Kollath_2010,delaCruz(2020)}.

In what follows, a spectral analysis is presented by exact diagonalization, based on averages over 40 numerical random realizations of the phase $\phi$ in the range $[0, 2 \pi]$\cite{Zhang}.

A widely employed signature of quantum chaos relates properties of the energy spectrum with those of random matrices
\cite{Stockman_2000}. Given an ordered set of eigenenergies $E_n$, with $n = 1, 2,...,d$, the nearest-neighbor spacing is given by $s_n = E_{n+1}-E_n$. Quantum chaotic systems have level spacing described by the Wigner-Dyson distribution, characterized  by level repulsion. In the Gaussian Orthogonal Ensamble (GOE) it reads
\begin{equation}
    P(\tilde s)= \frac{\tilde s}{2\sigma^2} e^{-\tilde s^2/4\sigma^2},
\end{equation}
in terms of the unfolded variable $\tilde s = s/\Delta E$, being $1/\Delta E$ the  density of states.

On the other hand, in regular systems, with non-correlated eigenvalues, the distribution of level spacing is well described by the Poisson distribution.
\begin{equation} 
p(\tilde s) = e^{-\tilde s}.
\end{equation}

To avoid the complication of the unfolding, in this work we employ the ratio of two consecutive level spacings proposed by Oganesyan and Huse\cite{Oganesyan_2007,Chavda_2014,Herrera_2020}.
\begin{equation}\label{ratio1}
r_n=\frac{\textrm{min}(s_n,s_{n-1})}{\textrm{max}(s,s_{n-1})}
\end{equation}
This quantity is independent of the  density of states and no unfolding is necessary. 
For quantum chaotic systems, the distribution of the level spacing ratio has an average of \cite{Atas_2013,Srivastava_2018}
\begin{equation}\label{ratio5}
\tilde{r}_W 
=4-2\sqrt{3} \approx 0.535,
\end{equation}
while for regular systems, it has a lower value of
\begin{equation}\label{ratio6}
\tilde{r}_P = 2 \ln 2 -1 \approx 0.386.
\end{equation}

\subsection{The disorder parameter and chaos}

\begin{figure}
    \centering
    \includegraphics[width=0.99\textwidth]{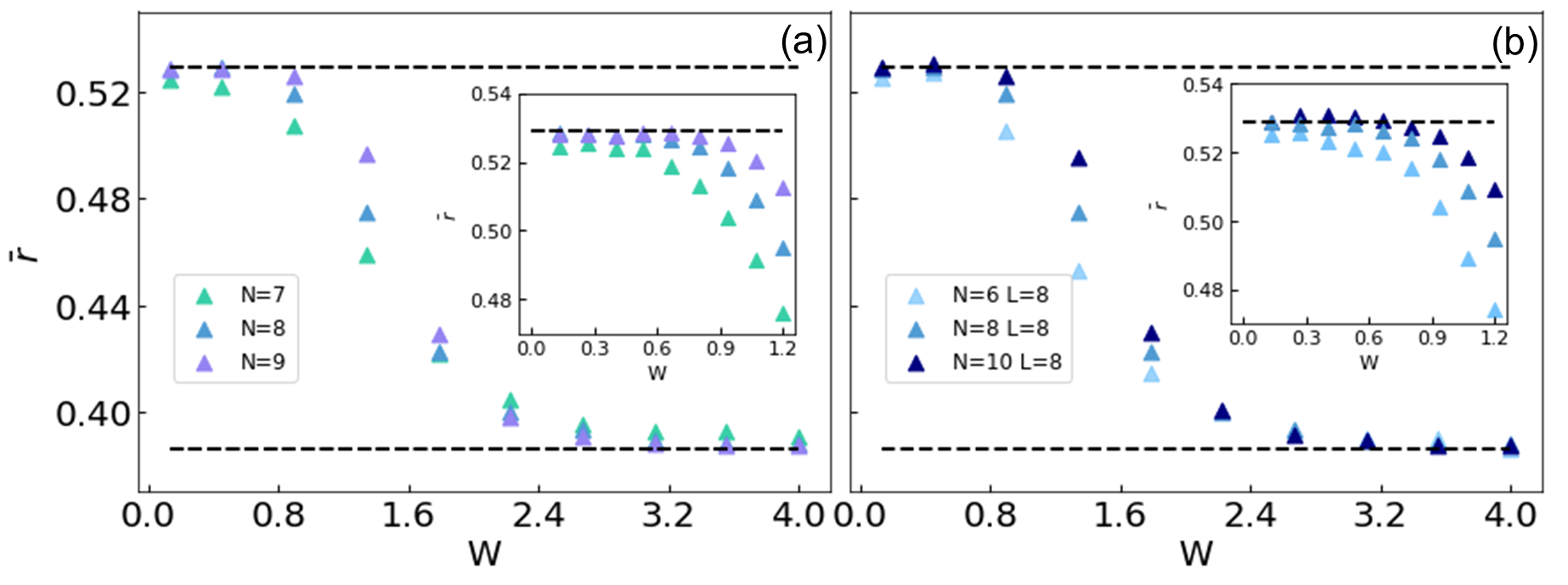}
\caption{ Average of two consecutive level spacing ratios as function of the disorder parameter $W$  for (a) $N=L \in [7,8,9]$ and  $W \in [0,4]$, (b) for different particle densities, $N= L \pm 2$,   $L= 8$ . Insets, zoom for $W \in [0,1.2]$. Only the 80 \% of the central part of the spectra was considered, by   neglecting the 10 \% lowest and 10 \% highest energy levels of the spectra. 
The upper and lower dashed lines show the limits
$\tilde{r}=0.53$, and $\tilde{r}=0.38$.}\label{fig:ratios88}
\end{figure}

The level spacing  ratio of Hamiltonian (\ref{eq:Hamiltonian})  is depicted in Fig. \ref{fig:ratios88} as a function of the disorder parameter $W$ for $N=L=[7,8,9]$ bosons and sites, and also for $L=8$ sites and $N=L\pm 2$ bosons, in a linear chain with open boundary conditions. It is clear that chaos is present in the region $W\leq 1$ for all the cases considered. A closer view  of this region in the inset shows that the presence of chaos remains at the same interval of $W$ irrespective of the number of sites and density. In this paper, the value $W=0.6$, at the middle of this region, is selected to study the chaotic dynamics. It is worth mentioning that we have excluded the case $W=0$ from the figures, since in this case the Bose-Hubbard model presents invariant subspaces that require a differentiated analysis of the level-spacing ratios in each subspace \cite{delaCruz(2020)}.

\subsection{Chaos and regularity in presence of weak disorder}

\begin{figure}
\centering
\begin{subfigure}
\centering
\subfigure{\includegraphics[width=0.7\textwidth ]{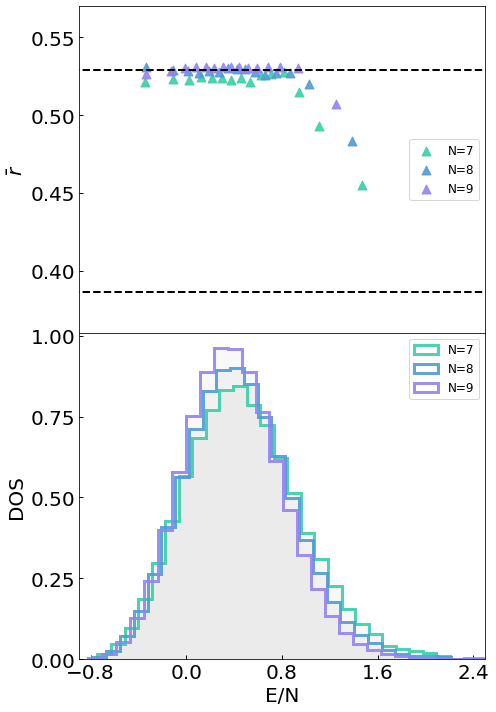}}
\end{subfigure}
\caption{(a) The level spacing ratio is shown for different   sets of the same number of consecutive states, for $W=0.6$. 
The upper and lower dashed lines show the limits $\tilde{r}=0.53$, and
$\tilde{r}=0.38$. (b) In shaded gray the DOS histogram is shown. Both figures represent $N=L \in [7,8,9]$. 
} \label{fig:ratio88_prom}
\end{figure}

In Fig. \ref{fig:ratio88_prom}~(a) a refined analysis of regularity and chaos is presented, employing the level spacing  ratio for the case of weak disorder $W=0.6$. Averages are made over 40 realizations of the Hamiltonian with different, randomly selected, disorder phases $\phi$, allowing to have enough statistics (around 500 energy level spacings) in small energy intervals centered at different energies. The dots represents the value $\tilde{r}$ 
for the energy levels in the intervals plotted versus the mean energy of the levels in those intervals. The figure shows that the chaotic region in the energy spectrum lies approximately in the interval $E/N=[-0.4,1.0]$.  As expected, at upper energy regions the quantum dynamics are regular. In Fig. \ref{fig:ratio88_prom}~(b),  the histogram of the density of states (DOS) is presented for the three different system sizes considered, showing a similar behaviour in terms of the scaled energy.

\section{\label{sec:initial_states} Classification of the occupation states and Participation Ratio
}

As mentioned before, in Ref. \cite{Lukin_2019} evidence of ETH was observed for weak values of the disorder parameter. In that reference the authors  considered  as initial state one atom in each site (the Mott state), which is a  particular state of the occupation or Fock  basis. In order to address possible deviations of this behaviour for other experimentally accessible  initial states, we consider in this study the whole set elements in the Fock basis as initial states. This set covers an ample range of energies, both in the chaotic  and regular regime of the spectrum and, as  shown below it presents a great diversity of behaviours which make it an adequate set to sample a generic initial state, even those with an initial entanglement between sites. This latter statement comes from the fact that the chaotic properties of the model makes that non initially entangled states, as those of the Fock basis, become entangled after a transient temporal interval.

The Fock  basis is defined as
\begin{equation}
    \ket k =    \ket{n_1, n_2, ...,n_L},
\end{equation}
with $L$ the number of sites, and 
\begin{equation}
    \sum_{i=1}^L n_i = N,
\end{equation}
where $N$ is the total number of atoms in the system. In this work we also use weak values of the disorder parameter ($W=0.6$) for three different sizes $N=L \in [7,8,9]$, allowing us to   perform a scaling analysis of our results. 

\begin{figure}
\includegraphics[width=1\textwidth ]{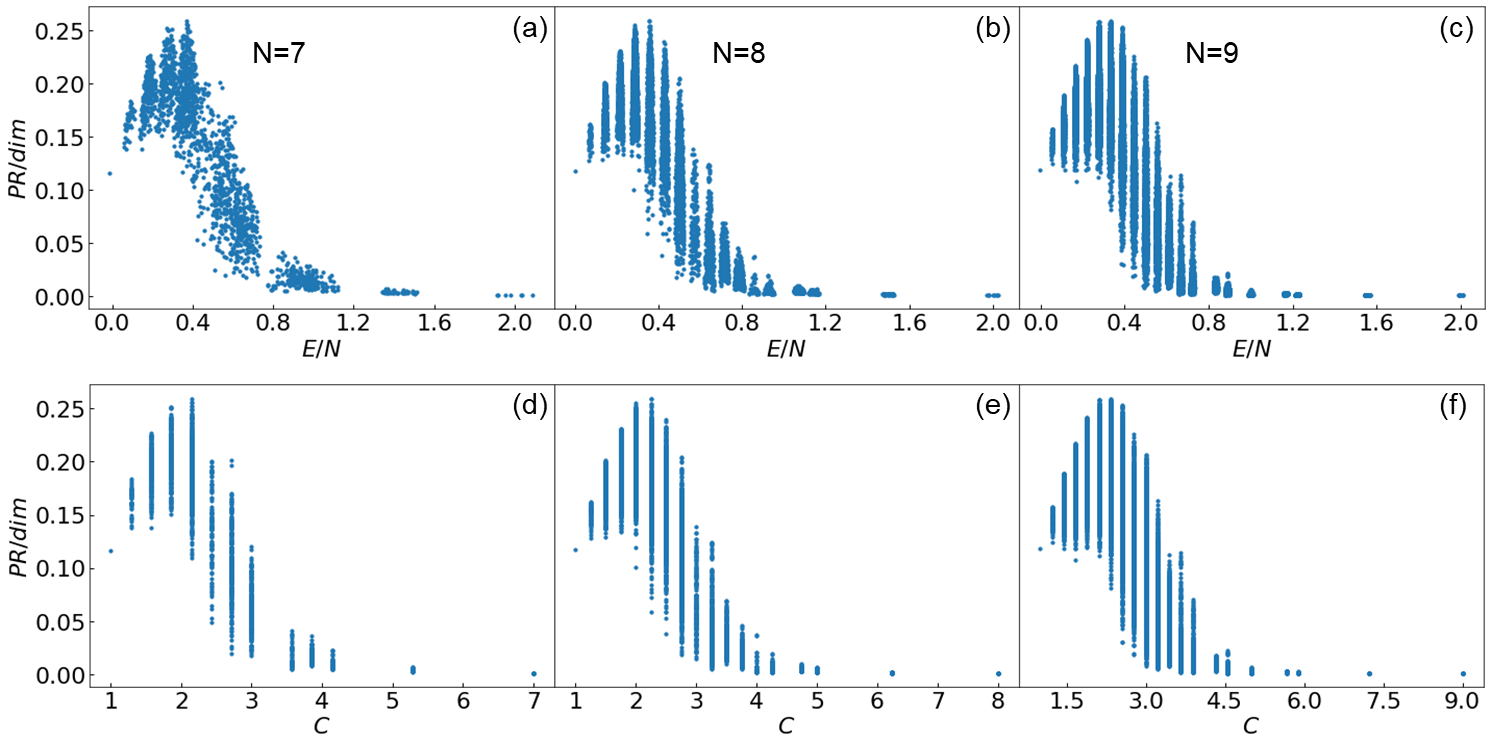}
\caption{(a)-(c) $PR/dim$ vs $E/N$ for all occupation states, (d)-(f) $PR/dim$ vs $C$, each column corresponds to different scale from $N=7$ to $N=9$.}\label{fig:prs_c}
\end{figure}

 In order to classify the states of the Fock basis and select those more adequate for our study, we consider several quantities. The first one is their mean energy per particle
 $E_k/N= \langle k |H|k\rangle/N$. The second quantity used in this classification is the Participation Ratio (PR) respect to the Hamiltonian eigenbasis. The PR provides an effective estimation of the number of states of a given basis $\{ \ket m \}$ (the Hamiltonian eigenbasis in this case), which are needed to built a given state $\ket k = \sum_m c^k_m \ket m$. It is defined as
\begin{equation}
\mathrm{PR(k)}= \frac {1} {\sum_m |c^k_m|^4} .
\label{PR}
\end{equation}
The largest the PR, the more delocalized is the state in this basis.
In  first row of Fig.~\ref{fig:prs_c} we show the PR divided by the dimension (dim) of the respective Hilbert space versus $E_k/N$ for all the states of the Fock basis in the three system sizes considered. While the eigenenergies range between $E/N=-0.8$ and $E/N=2.4$, the mean  energies per particle  of the
occupation states are more concentrated, with most of them having energies
between 0 and 1. Note that most of the states in the occupation basis have
mean  energies in the middle of the spectrum which is chaotic according to Fig.\ref{fig:ratio88_prom}. Additionally, observe that the states are organized in several clusters with approximately the same energy. The cluster with lowest energy consists of only one state, the Mott state with one particle per site.  The different clusters can be labeled  by the {\it crowding parameter} 
\begin{equation}\label{crouwd}
C=\frac{1}{N}\sum{n_i^2},    
\end{equation}
which ranges from $1$, for the Mott state to $N$ when all the particles are located in only one site.
Observe that, since the hopping term in (\ref{eq:Hamiltonian}) has no diagonal components in the occupation  basis, and the disorder term is an oscillatory function of the phase $\phi$ with null average over many realizations, the average energy of the occupation states has only contributions from  the interaction term [the second one in the Hamiltonian of Eq. (\ref{eq:Hamiltonian})]. For this reason, the average value of the energy is approximately linear in the crowding parameter $E_k/N= U(C-1)/2$.

Second row of Fig.~\ref{fig:prs_c} presents the same ratio PR/dim of all occupation states  plotted against their crowding parameter. Note that the clusters of energy become vertical lines, facilitating the classification of the states.  
It can be seen that most of the states, whose $C$ lie between $1$ and $2.5$ have the largest PR and $E_k/N<1$. On the other hand, those states with larger $C$ have  highest energies and lowest PR values. Therefore, in general terms, states with small $C$ are less localized in the Hamiltonian eigenbasis than those with larger $C$,  which are far more localized. 
Nevertheless, observe that the states with small $C$ have  a considerable dispersion in their PR values. For instance, for states with $C\approx 2$ the ratio PR/dim ranges approximately from 0.1 up to 0.25.

In the next section we study the dynamics of initial Fock states with a crowding parameter $C\sim 2$ which have a mean energy per particle well inside the chaotic region of the spectrum. We consider the  three different system sizes and for a given $C$ we select states with high and low PR in order to detect differences in their dynamics. The dynamical quantity we choose to perform our study is the survival probability, whose definition and properties are discussed in the  following section.

\section{The Survival probability and the correlation hole \label{sec:survival_probability}}

The survival probability $S_P(t)$ is defined as the probability to find any initial state $\ket{\Psi(0)}$ after an evolution with time $t$,
\begin{equation}
    S_P(t) = \left|\braket{\Psi(0)|\Psi(t)}\right|^2 =  \left|\sum_m |c_m|^2 e^{-iE_m t} \right|^2,
\label{eq:SP1}    
\end{equation}
$c_m$ are the components of the initial state  in the Hamiltonian eigenbasis, $|\Psi(0)\rangle=\sum_m c_m |E_m\rangle$.
The survival probability is closely related to the spectral form factor (SFF) \cite{Brezin1997}, which is the survival probability of an initial state whose components are given by a Gibbs thermal distribution $|c_m|^2= \frac{e^{-\beta E_m}}{Z(\beta)}$ with $Z(\beta)=\sum_m e^{-\beta E_m}=\mathrm{Tr} \ e^{-\beta H}$.

The survival probability (SP)  has been used in recent works as a quantum indicator of  chaos, partly because for random initial states there exists an analytical expression that relates the two-level form factor of GOE full random matrices with the dynamics of the survival probability at long times \cite{Alhassid(1992),Torres-Herrera(2017), Torres-Herrera(2017-2),Torres(2018),Torres-Herrera(2019), Schiulaz(2019),Lerma-Hernandez(2019), delaCruz(2020)}. This implies that at long temporal scales  the behavior of this observable is universal, i.e. it is dictated by the same spectral correlations as those  of  gaussian ensembles of random matrices.  From an experimental point of view, the SP can detect chaos  without the need and  complications of measuring the  complete set of eigenenergies. Additionally, the SP  can detect quantum chaos even
if there are some hidden symmetries \cite{delaCruz(2020)}, which is a limitation  of nearest-neighbor spacing distributions that   are strongly affected by hidden symmetries in the models.
 
 If there are no degeneracies\cite{delaCruz(2020), Giraud2022}, the temporal average of the survival probability at long times approaches the Inverse Participation Ratio 
\begin{equation}
\braket{S_P^\infty}=\mathrm{IPR}=\sum_m |c_m|^4.
\end{equation}
It is the inverse of the PR of Eq. \ref{PR}, which provides an estimation of the number of elements of a given basis (the energy eigenbasis, in this case) participating in an arbitrary quantum state.

Random matrix theory allows to obtain an analytical expression for the survival probability of an ensemble of random states with a given  smoothed local density of states $\rho(E)$ (see below) and energy components in the  chaotic region of the spectrum characterized by level correlations of the  Gaussian orthogonal ensemble (GOE). It reads \cite{Lerma-Hernandez(2019),Villasenor2020,delaCruz(2020), Alhassid(1992), Torres-Herrera(2017), Torres-Herrera(2017-2), Torres-Herrera(2019)}
\begin{equation}
    \braket{S_P(t)} = \frac{1-\mathrm{IPR}}{\eta -1}\left[\eta S_P^{bc}(t) - b_2\left(\frac{t}{2 \pi \bar\nu}\right)\right] + \mathrm{IPR},
    \label{eq:S_p-analitical}
\end{equation}
where $S_P^{bc}(t)$ is given by the  Fourier transform of the smoothed Local Density of States as
\begin{equation}
S_P^{bc}=\left|\int \rho(E) e^{-i E t} dE \right|^2,
\label{eq:initialSP}
\end{equation}
the mean density of states $\bar{\nu}$ is taken in the energy region where the components  $c_m$ are more important, and the parameter  $\eta$ takes into account the avalaibility of energy levels,


\begin{equation}\label{eq:eta2}
\eta=\frac{1}{\int dE \frac{\rho^2(E)}{\nu(E)}}.
\end{equation}
The details of the   estimation of this parameter are explained in Appendix A.

\begin{figure}
\centering

\begin{subfigure}
\centering
\includegraphics[width=1\textwidth ]{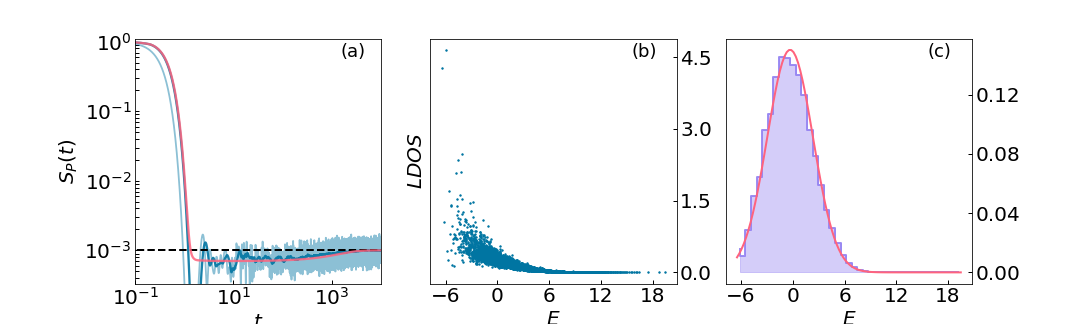}
\end{subfigure}
\caption{(a) Survival probability for  the Mott ( $N=L=8$) state $\ket{\Psi (0) }=\ket{1,1,..,1}$, (b) the LDoS ($\times 10 ^{-3}$) and (c) the smoothed  LDoS histogram for constant windows of energy $\Delta \mathcal{E}=0.74$ for only one realization in $\phi$, in red the Gaussian fit.
}\label{fig:LDOS}
\end{figure}

For a given state, the local density of states (LDoS) is defined as
\begin{equation}
\rho_{L}(E)=\sum_{m} |c_m|^2 \delta(E-E_m).
\label{ed:LDOSexact}
\end{equation}
A smoothed local density of states $\rho(E)$ can be  obtained from the components of the initial state  by fitting a continuous  distribution to the histogram 
$$
\rho(\mathcal{E}_i)=\!\!\!\!\!\!\!\!\!\sum_{\substack{ m \text{\ \ with} \\
|\mathcal{E}_i-E_m|<\Delta  \mathcal{E}/2} }\!\!\!\!\!\!\!\!\! |c_m|^2,
$$
for a set of equally spaced energy values  $\mathcal{E}_{i+1}-\mathcal{E}_i=\Delta \mathcal{E}$. The value $\Delta\mathcal{E}$ determines the degree of resolution with which $\rho(E)$ describes the LDoS, $\rho_L(E)$. For $\Delta\mathcal{E}$ less than the difference of consecutive levels we obtain maximal resolution, whereas for  $\Delta\mathcal{E}\sim \Delta H$ we obtain a smoothed $\rho(E)$ which determines the initial decay of the survival probability in Eq.\eqref{eq:S_p-analitical}.   In Fig.\ref{fig:LDOS}(c) we show this smoothing of the LDoS for a representative initial state, for which we fitted a  Gaussian distribution.  
The next term in Eq. \eqref{eq:S_p-analitical} is the two-level form factor of the GOE ensemble \citep{Alhassid(1992)},
\begin{equation}
        b_2(t) = [ 1-2t +t \ln (2t + 1) ] \Theta(1-t) + \left[t \ln \left( \frac{2t+1}{2t-1} \right) -1 \right] \Theta (t-1),
\end{equation}
where $\Theta$ is the Heaviside step function. The two-level form factor evolves the survival probability from a minimum to its relaxation value  $\braket{S_P^\infty}$, performing a dip known as the {\it correlation hole}. 
As mentioned above, for a given initial state in the occupation basis we consider 40 realizations of the Hamiltonian.  For each realization we calculate numerically the survival probability and, at each time, we consider its average over the realizations. By using   the Mott state, an  illustrative result of this procedure is shown by the light curve in Fig.\ref{fig:LDOS}(a).  We  see that, after the initial decay,  the averaged survival probability  displays temporal oscillations. In order to smooth these quantum fluctuations, we consider additionally a temporal rolling average, which is shown in the same figure by a dark blue line. This rolling temporal average makes more evident  the  presence of a hole in the survival probability,  described by Eq.~\eqref{eq:S_p-analitical} (red line  in the figure), which is a  direct signature of the existence of correlated eigenvalues.  The red line in  Fig.\ref{fig:LDOS}(a) was obtained by considering a Gaussian fit to the smoothed LDoS of Fig.\ref{fig:LDOS}(c). The LDoS at full resolution is shown in Fig.\ref{fig:LDOS}(b). It is worth mentioning that  the correlation hole  does not appear in systems with uncorrelated eigenvalues, which is not the case of the Mott initial state of Fig.\ref{fig:LDOS},

\subsection{Numerical results}
\label{sec:results}
\begin{figure}
\centering
\includegraphics[width=1\textwidth ]{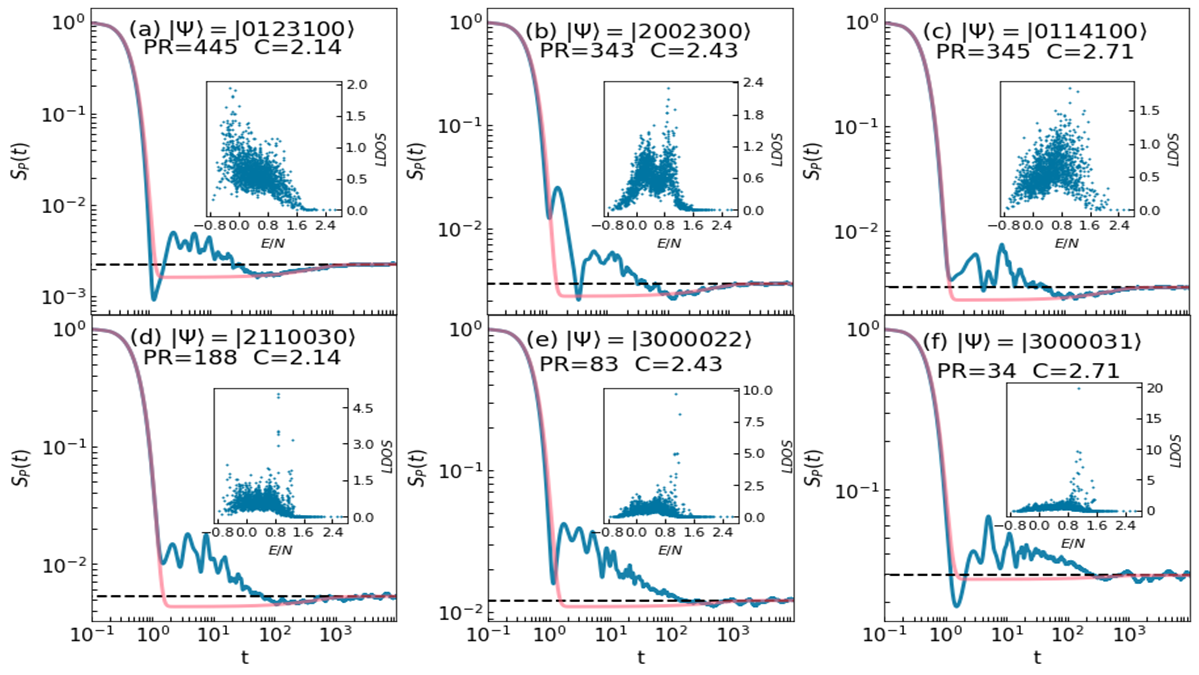}
\caption{Survival probability of selected states for $N=L=7$ with different crowding parameter between $C=[2,3)$. The states with the largest $PR$ value are in the first row((a), (b), (c), the states with the lowest PR are in the second row. 
The blue line is the smoothed survival probability averaged over 40 realizations of the disorder parameter, the analytic expression is shown as red line. Inside: the average LDoS of the corresponding state}\label{fig:SP_maxmin7}
\end{figure}

\begin{figure}
\centering
\includegraphics[width=1\textwidth ]{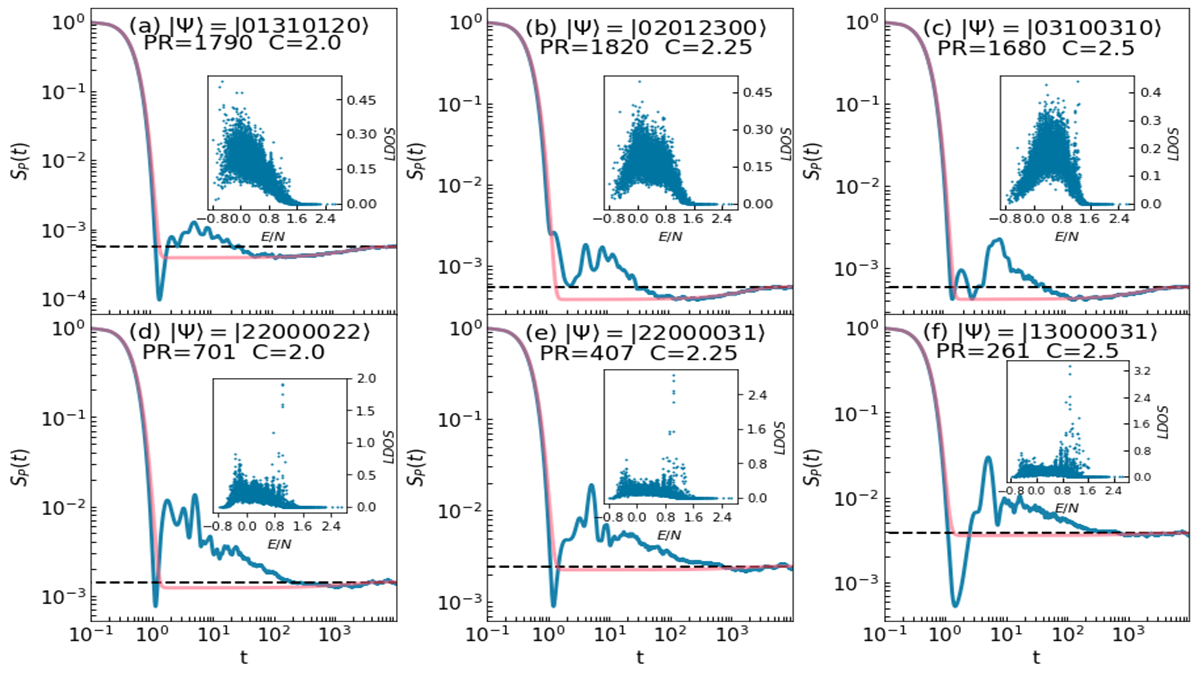}
\caption{Survival probability of selected states for $N=L=8$ 
 with different crowding parameter between $C=[2,3)$. The states with the largest $PR$ value are in the first row((a), (b), (c), the states with the lowest PR are in the second row.
The blue line is the smoothed survival probability averaged over 40 realizations of the disorder parameter, the analytic expression is shown as red line. Inside: the average LDoS of the corresponding state}\label{fig:SP_maxmin8}
\end{figure}

\begin{figure}
\centering
\includegraphics[width=1\textwidth ]{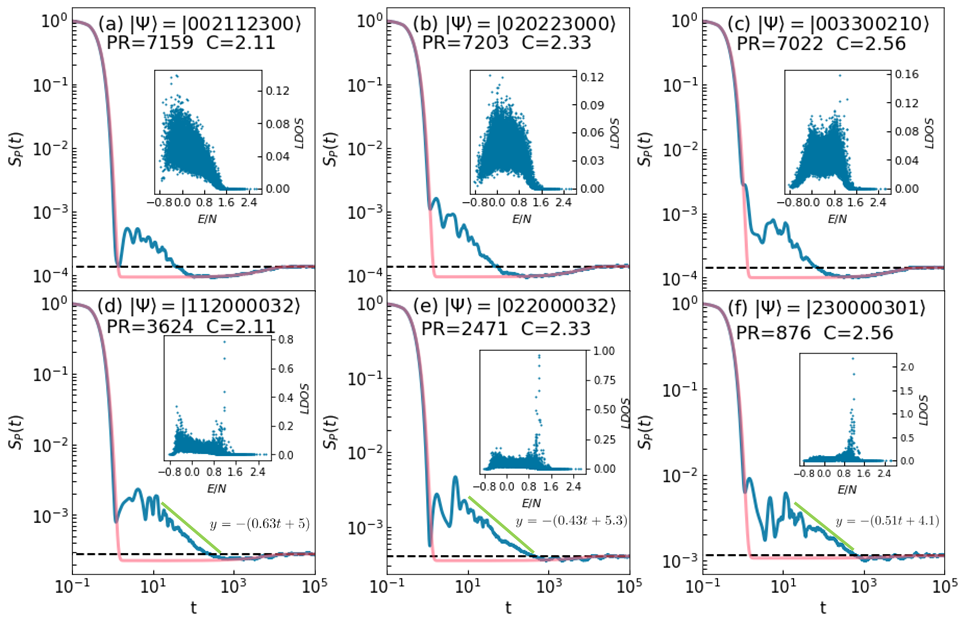}
\caption{Survival probability of selected states for $N=L=9$ with different crowding parameter between $C=[2,3)$. The states with the largest $PR$ value are in the first row (a), (b), (c), the states with the lowest PR are in the second row (d), (e), (f).
The blue line is the smoothed survival probability averaged over 40 realizations of the disorder parameter, the analytic expression is shown as red line. Inside: the average LDoS of the corresponding state. Green lines in the bottom panels depict the power-law decay of the survival probability, whose exponent is given by the slope of the linear fits also shown next to the lines.}\label{fig:SP_maxmin9}
\end{figure}

\begin{figure}
\centering
\includegraphics[width=1\textwidth ]{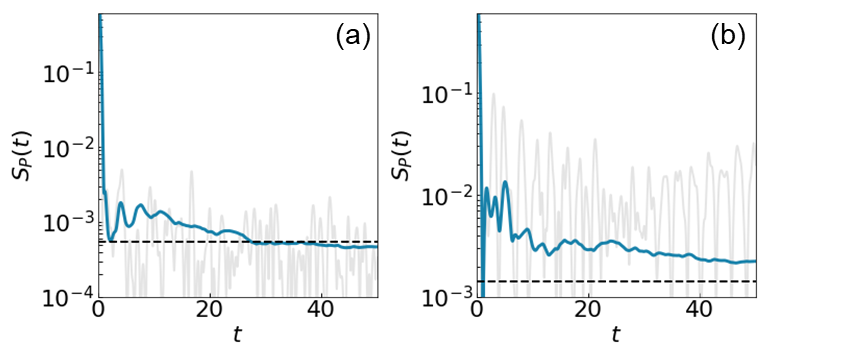}
\caption{(a) Survival probabilities for the 
state $\ket{\Psi (0) }=\ket{0,2,0,1,2,3,0,0}$ with high PR and (b) for a state 
$\ket{\Psi (0) }=\ket{2,2,0,0,0,0,2,2}$, with low PR. In both cases for  $N=L=8$. The gray line corresponds to one random realization in $\phi$ and the blue line is the average over 40 realizations. These two states are the same considered in Fig.\ref{fig:SP_maxmin8}(b) and (d), respectively.
}\label{fig:SPrevivals}
\end{figure}

We focus, for the three different system sizes, on initial states of the Fock basis with a mean energy inside the already identified chaotic region and study their survival probabilities.  For that, for each system size,  we select, as representative of the diverse behaviours of the whole basis, 6 initial  states of the Fock basis with three different values of  the crowding parameter  $C$  in the interval  $C=[2,3]$. For 
given system size  and  $C$,  we consider  a pair of  states, one with the highest PR and a second one  with the lowest PR (in  \ref{appx:2}, for $N=L=8$,  results for states with  intermediate values of the PR are presented). This  allows us to contrast the differences in the  behaviour of the SP.  Varying randomly $\phi$, we consider 40 realizations of the Hamiltonian for each initial state. We perform the same average over realizations and  the same rolling temporal average as in Fig.\ref{fig:LDOS}~(a) to obtain the   results  shown by blue lines  in Figs. \ref{fig:SP_maxmin7}, \ref{fig:SP_maxmin8}  and \ref{fig:SP_maxmin9} for $N=L=7,8,$ and $9$, respectively. In all these figures, we also plot with a red line the  analytical expression of  Eq.\eqref{eq:S_p-analitical}.  

In all cases  (Figs. \ref{fig:SP_maxmin7}, \ref{fig:SP_maxmin8}, and \ref{fig:SP_maxmin9}), the initial decay of the survival probability is Gaussian and adequately described  by the red lines obtained from the  squared magnitude of the Fourier transform, eq.\eqref{eq:initialSP}, of the smoothed LDoS which is also Gaussian. Likewise, in all cases the asymptotic value of the SP is given by the IPR, which is attained at very long times. Important differences in the behaviour of the SP can be observed at intermediate temporal scales. 

After the initial Gaussian decay, 
the survival probability of all the states  display a non-generic behavior, with oscillations whose amplitudes seem to decay according to a power-law \cite{Tavora2016} (see green lines in bottom panels of Fig. \ref{fig:SP_maxmin9}). At this scale, the analytical red lines do not describe, not even approximately, the numerical results. The origin of this behavior can be traced back  to the details of the components of the initial state in  the eigenbasis, the LDoS, whose  averages  over the 40 realizations are shown as insets in all the panels of the figures.

The main differences in the behaviour of the SP comes after the end of this power-law-decay regime.
For states with large $PR$ (upper row in  figures \ref{fig:SP_maxmin7}, \ref{fig:SP_maxmin8}, and \ref{fig:SP_maxmin9}), the power-law regime  ends at the bottom of the correlation hole that is clearly visible in the numerical (blue) lines and analytical (red) lines. From this time up to equilibration the numerical SP is very well described by the analytical expression and  
behaves according to the GOE correlations in the spectrum.  For these states with large PR,  the numerical SPs exhibit a  ramp towards the asymptotic value that  corresponds to the final stage of the correlation hole. Correspondingly,  for these high-PR states  we can observe a very dense  LDoS  in which all the Hamiltonian eigenstates in the chaotic region ($E/N\in[-0.4,1]$) participate. 
  The correlation hole ends at $t\approx 10^{3}$ for $N=7$, at $t\approx 10^{3.5}$ for $N=8$,  and for $N=9$ it ends at $t\approx 10^4$, which is approximately the  corresponding Heisenberg time for each case, $t_H\propto\frac{\textrm{dim}}{N} $.

On the other hand, for states with low PR (bottom row in all figures 
\ref{fig:SP_maxmin7}, \ref{fig:SP_maxmin8}  and \ref{fig:SP_maxmin9}) 
 the survival probability 
 behave very differently respect to  the states with  high PR, although all of them  are in the chaotic  region where ETH and RMT is valid.
There are two main features in the survival probability  differentiating their behaviour.
The first one is that, for low-PR states,   the power-law-decay regime persists for a longer time and ends at a time  approximately one order of magnitude longer than in the states with large PR. The second one is that, for low-PR states, this power-law regime ends directly at the asymptotic value and a barely identifiable or even absent  correlation hole is obtained for the SP of these states. The presence of the power law decay is illustrated  in Figs. 7 (d,e,f), where a diagonal green straight line is plotted parallel to the peaks of the SP, with a slope $0.5 \pm  0.1$. It means that in this temporal region, the survival probability of these states decays approximately as $t^{1/2}$.

The averaged LDoS display  differences as well. In the insets of the panels for states with low PR (lower row in the figures), it can be observed that a small number of eigenstates have very large components, which explains the anomalous behaviour of the survival probability and the low PR of these states. 
It is worth noting the similarities of these low-PR states with the so-called many-body scarred states reported in other models \cite{Serbyn_2021,chandran2023quantum,turner2018quantum,Lin2019,Hudomal2020,hummel2023genuine,dooley2020enhancing,su2023observation}.
The scarred states are characterized by displaying persistent, long lasting revivals, the participation of a  small set  of eigenstates, which keep them trapped in a small, low entropy subspace which do not thermalize \cite{su2023observation}. As a consequence, they do not display a correlation hole in their survival probability.
In this model, another interesting property  of these low-PR states 
is that  initially the  bosons are more concentrated near the borders of the linear chain,  with  the middle sites  empty. This suggests a connection between persisting revivals and boundary effects. A comparative study between the case studied here with the case of periodic boundary conditions would permit to establish this possible connection.

Finally, in order to unveil the properties of the SP that produce the power-law decay obtained in the ensemble average of the SP, both for states with high and low PR, in Fig.~\ref{fig:SPrevivals} we show the SP of two representative initial states, one with high PR and another one with low PR. Differently to previous figures, here we show the SP for only one realization of the Hamiltonian and use a linear scale in the horizontal time axis. We can see that in the case of the high-PR state we obtain revivals appearing randomly, whereas  for the state with low PR the revivals are periodic and persists for a longer time, reinforcing the ideas that   these low-PR states are scarred states.

\section{\label{sec:conclusion}Conclusions}

We have presented a detailed study of the dynamics of  states in the occupation basis under unitary evolution  in the   Interacting Aubry-Andr\'e model, which has been recently realized experimentally, showing  evidence of thermalization  for the Mott state, which is a particular element of the occupations basis. 
For different  number of particles ($N$) and sites ($L$), we have studied the spectral signatures of chaos, confirming that the system becomes chaotic 
when the disorder parameter $W$ is small compared with the other Hamiltonian parameters. 
 The system exhibits chaos in an intermediate  energy per particle interval, 
  being regular in the low and high energy extreme of the spectrum. We have verified that this energy per particle  interval  is approximately the same    for the three different system sizes considered $N=L=$7, 8, and 9, which have Hilbert space dimensions varying in one order of magnitude, from $1.7\times 10^3$ up to $2.4\times 10^4$. 

We have identified states of the occupation basis that show an anomalous behaviour in their dynamics.  These states show lasting revivals and, although their energy components are located in the already identified chaotic region of the model, they do not show  in their dynamics  signatures of the correlated   energy spectrum. These states have a low  participation ratio of the energy components. On the contrary, states with large values of the participation ratio manifest the random-matrix correlations of the energy spectrum in  their dynamics, in the form of a correlation-hole in their survival probability.

The states with low participation ratio and anomalous behaviour in their dynamics have similar characteristics as the so-called quantum scarred states, namely, lasting revivals and absence of a correlation-hole in the dynamics of their survival probability.  

Additionally, we identified the closeness of the initial atoms to the border of the chain of sites as the origin of occupation states with small PR and anomalous dynamics.  
A comparison of the behaviour of the  open chain  studied here with a chain with periodic conditions would shed light about the possible relation between states with persisting revivals and border effects.  

{\bf Acknowledgments}: We acknowledge the support of the Computation Center - ICN, in particular to Enrique Palacios, Luciano DÃ­az, and Eduardo Murrieta. This research was funded in part by the DGAPA- UNAM project number IN109523.

{\bf Competing interests}: the authors have no competing interests to declare.

{\bf Credit roles}: 
{\bf Carlos Diaz Mej\'ia:} Formal analysis, Software, Visualization, Writing - original draft,
{\bf Javier de la Cruz:} Formal analysis, Software, Visualization,

{\bf Sergio Lerma-Hern\'andez:} Conceptualization, Formal analysis, Writing - review \& editing,
{\bf Jorge G. Hirsch:} Conceptualization, Formal analysis, Writing - review \& editing.

\appendix
\section{Effective dimension and survival probability at short times}\label{appx:1}

One of the parameters of the survival probability is the effective dimension $\eta$. 
The numerical estimation of $\eta$ requires an integral involving the LDoS of a given state and the DOS of the whole system,
\begin{equation}
\eta=\frac{1}{\int dE \frac{\rho^2(E)}{\nu(E)}}.
\label{eq:eta}
\end{equation}

We explored performing the above integral in two different ways: approximating the Dirac deltas in the   LDoS, $\rho_L(E)$,  by Gaussian distributions of width $\Delta E$, or performing  a Riemann Sum with energy intervals given by $\Delta E$.
We found that the Riemann Sum shows less dispersion in the estimation of the $\eta$ parameter.

\begin{figure}
     \begin{subfigure}
         \centering
         \subfigure{\includegraphics[width=0.48\textwidth ]{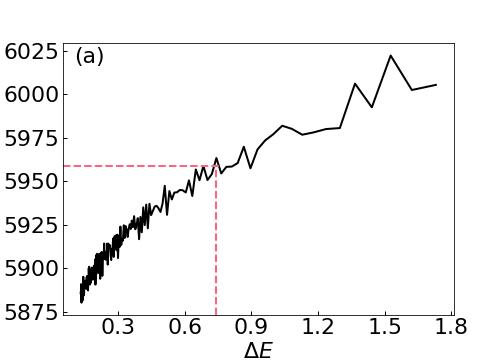}}
     \end{subfigure}
     \centering
     \begin{subfigure}
         \centering
         \subfigure{\includegraphics[width=0.48\textwidth ]{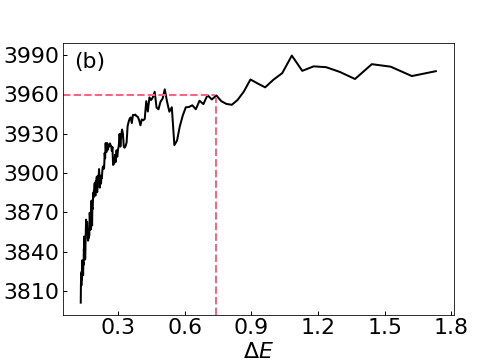}}
     \end{subfigure}
\caption{$\eta$ vs $\Delta E$ for the state (a) $\ket{\Psi(0)}=\ket{022000220}$ and (b) $\ket{\Psi(0)}=\ket{11111111}$ for  40 realizations in $\phi$, the red lines are the values chosen in the numerical calculation, both corresponds to a $\Delta E =0.74$}\label{fig:eta_deltae}
\end{figure}

To select the most convenient energy interval $\Delta E$ for performing the integral as a Riemman sum, we calculated $\eta$ for $\Delta E \in [0.2, 1.8]$ for the state (a) $\ket{\Psi(0)}=\ket{022000220}$ and (b) $\ket{\Psi(0)}=\ket{11111111}$ considering   40 realizations in $\phi$. For the   first state, we obtain a value of  $\eta=5960 \pm 85$, whereas for  the second state  we obtained  $3960 \pm 30$, for $\Delta E \in  [0.3,1.8]$. These average values are shown as red lines in Fig. \ref{fig:eta_deltae}.  Both averages   correspond  to a  $\Delta E =0.74$. In this way, we are confident that the selected value of $\eta$ would not vary in more than 2\% if we would have employed a different energy interval $\Delta E$.

At short times, when the revivals have vanished and the correlation hole starts, the averaged survival probability has its lowest value along the ramp which will end at long times in the IPR. This bottom value is \cite{Lerma-Hernandez(2019)} 

\begin{equation}
SP_{low} \approx IPR - \frac{1}{\eta}.
\end{equation}

This expression shows that the maximum depth of the correlation hole, below its asymptotic value, is  $\frac{1}{\eta}$.

\begin{figure}
     \begin{subfigure}
         \centering
         \subfigure{\includegraphics[width=0.8\textwidth ]{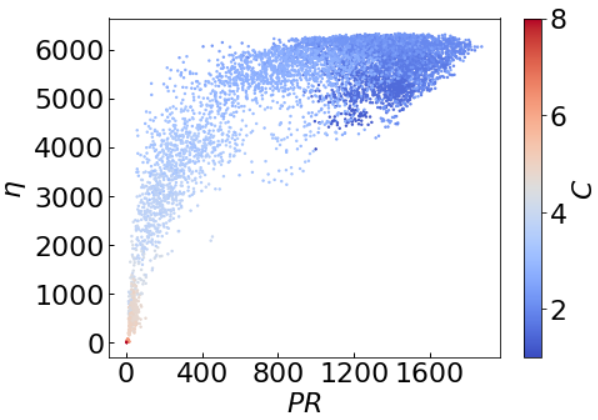}}
     \end{subfigure}
\caption{ $\eta$ vs $PR$ of each initial state vs $C$  for 40 realizations in $\phi$}\label{fig:eta_pr}
\end{figure}

As the maximum depth of the correlation hole is $1/\eta$ below the asymptotic value $IPR = 1/PR$, it is interesting to know the relationship between these two quantities. In Fig. \ref{fig:eta_pr} we show a plot of $\eta$ vs $PR$, with the color code depicting the value of $C$, for all the states in the occupation basis, for a system of $N=8$ bosons in $L=8$ sites. For those states with PR values  smaller than 400, which have the largest values of C and  highest energies, $\eta$ grows linearly with PR, with a relation close to $\eta \approx 10 PR$. For  states in the  low energy and chaotic region, $C \leq 4$, most of the states with $PR > 800$ have a nearly constant value $5000 <\eta < 6000$.  

\section{Emerging of the correlation hole.}\label{appx:2}
\begin{figure}
     \begin{subfigure}
         \centering
         \subfigure{\includegraphics[width=1\textwidth ]{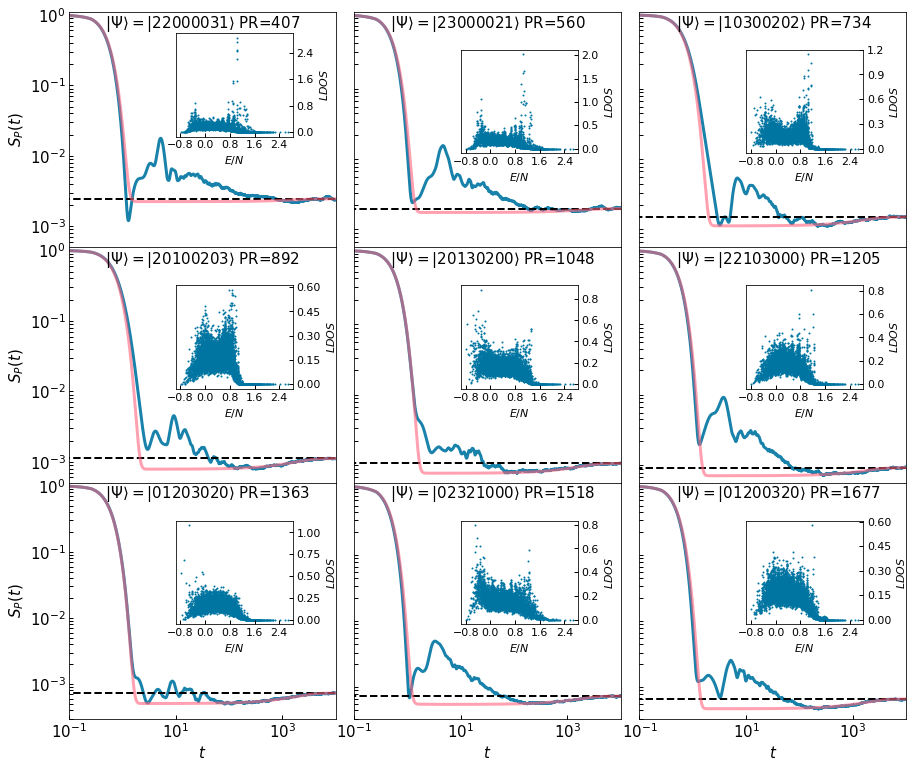}}
     \end{subfigure}
\caption{ Set of survival probabilities for  $N=L=8$ and $C=2.25$. They are shown from its lowest $PR=407$ up to $PR=1677$ at constant steps. Inset: averaged LDoS of the corresponding state}\label{fig:sp_appx3}
\end{figure}

\begin{figure}
     \begin{subfigure}
         \centering
         \subfigure{\includegraphics[width=1\textwidth ]{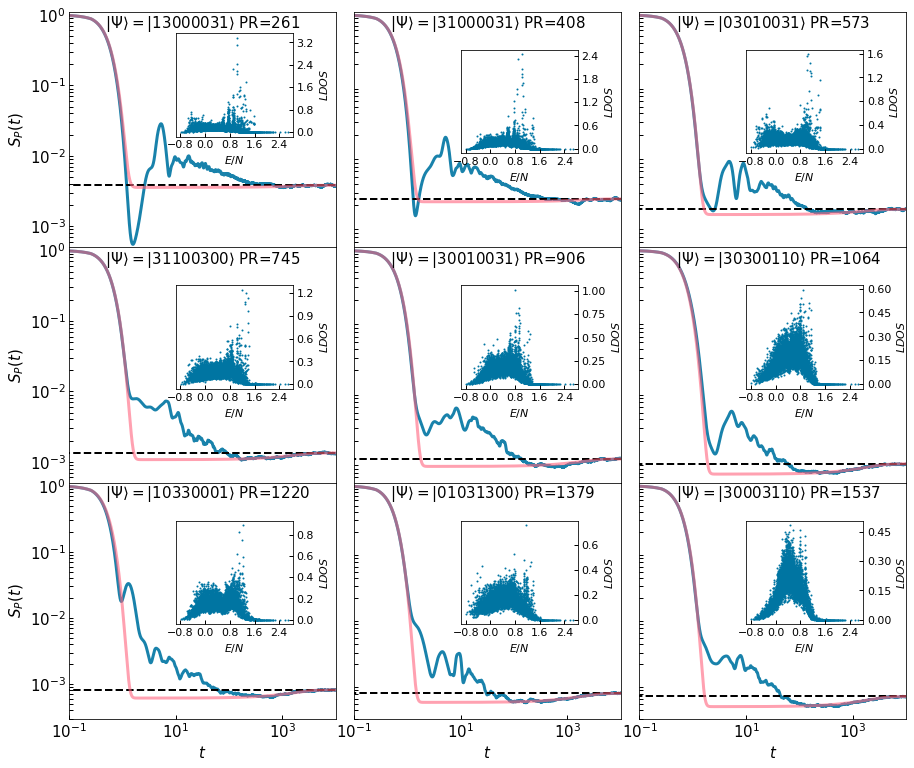}}
     \end{subfigure}
\caption{ Set of survival probabilities for  $N=L=8$ and $C=2.5$. They are shown from its lowest $PR=261$ up to $PR=1537$ at constant steps. Inset: averaged LDoS of the corresponding state }\label{fig:sp_appx4}
\end{figure}

In this section we describe how the correlation hole starts to appear for a specific value of the crowding parameter in the case $N=L=8$. Figures \ref{fig:sp_appx3} ($C=2.25$) and \ref{fig:sp_appx4} ($C=2.5$) displays how the survival probability evolves as  the $PR$ increases.

We can observe that states with the lowest $PR$ have no correlation hole and their SP shows  a polynomial decay that persists until attaining its equilibration value. Their LDoS show as well shows   large concentration in few eigenstates. In states with increasing PR the LDoS starts to be distributed along  the eigenstates and the correlation hole emerges. It is noticeable that only a small set of states lacks of correlation hole, and that  for these states the  bosons  tend to be on the borders of the chain.

\clearpage
\newpage

\bibliographystyle{unsrt} 

\end{document}